\documentclass[aps,nofootinbib,prd,eqsecnum,showpacs,showkeys,preprintnumbers]{revtex4-2}
\usepackage[caption=false]{subfig}
\usepackage{graphicx}
\usepackage{amsmath}
\usepackage{amsfonts}
\usepackage{amssymb}
\usepackage{float}
\usepackage{color}
\usepackage{bm}
\usepackage{mathrsfs}
\usepackage{epstopdf}
\usepackage{url}
\usepackage{footnote}
\usepackage{textcomp}
\usepackage{ulem}
\usepackage{esint}
\usepackage[unicode=true, pdfusetitle,
 bookmarks=true,bookmarksnumbered=false,bookmarksopen=false,
 breaklinks=false,pdfborder={0 0 1},backref=false,colorlinks=false]{hyperref}
\usepackage{multirow}
\usepackage{pifont}

\begin{document}

\title{Exploring the impact of Tree Level Higgs Potential on Reheating in $%
f(Q)$ Gravity}
\author{K. El Bourakadi$^{1,2}$}
\email{k.elbourakadi@yahoo.com}
\date{\today }

\begin{abstract}
This article investigates the effects of Tree Level Higgs Potential on
reheating in $f(Q)$ gravity. We examine the formalism of tree level
inflation in a modified $f(Q)$ model, finding that the slow roll and e-folds
number during inflation depend on the $f(Q)$ model $\alpha $ parameter.
Additionally, we calculate the reheating as a function of the decay rate,
constraining the inflationary parameters and reheating temperature. Our
study reveals that the Tree Level Higgs Potential influences the reheating
temperature and decay rate. We conclude that increasing the decay rate
energy leads to higher reheating temperature, and the chosen $\alpha $
parameter affects the reheating temperature curve. This research advances
our understanding of the interplay between Tree Level Higgs inflation,
matter production, and modified gravity in $f(Q)$ gravity.
\end{abstract}

\affiliation{$^{1}${\small Quantum Physics and Magnetism Team, LPMC, Faculty of Science
Ben M'sik,}\\
{\small Casablanca Hassan II University, Morocco}\\} 
\affiliation{$^{2}$LPHE-MS Laboratory department
of Physics, Faculty of Science, Mohammed V University in Rabat, Morocco.\\}

\maketitle

\section{Introduction}

\label{sec1}

%%%%%%%%%%%%%%%%%%%%%%%%%%%%%%%%%%%%%%%%%%%%%%

The most likely successful inflation scenario is the single-field chaotic
model \cite{I1,I2}. However, polynomial and monomial potentials are also
viable options. For example, many monomial potentials have quadratic or
quartic forms, like the simple double-well potential used to explain the
production of Primordial Black Holes \cite{I3,I311,I32}. Additionally, Ref. 
\cite{I4} employed the Higgs potential at the tree level to develop
inflation in non-supersymmetric grand unified theories $\left( GUT\right) $
and calculate proton lifetime. Similarly, Refs. \cite{I5} and \cite{I6}
explored the effects of inflation couplings with other particles, such as $%
\left( GUT\right) $ scalars, to investigate quantum smearing. They found
that the Higgs potential can be divided into three branches. Inflationary
models were created to resolve several problems associated with Big Bang
cosmology, including the flatness problem, horizon problem, homogeneity, and
the numerical density of monopoles \cite{I7}. Additionally, inflation
provides a causal explanation for the observed anisotropy of the cosmic
microwave background radiation $\left( CMB\right) $ \cite{I8}. Recently,
braneworld models have emerged as a novel alternative approach to test
certain predictions and incorporate corrections to general relativity \cite%
{I9,I10}.

Within the framework of GR, the motion of spacetime is determined by the
Ricci curvature scalar $R$, which is derived from the Levi-Civita
connection. Gravity is considered a metric theory in this context, meaning
it adheres to established assumptions such as Lorentz invariance and
causality. However, there are other geometric quantities such as torsion and
non-metricity that can also be used to describe the gravitational field.
This has been explored in previous studies \cite{I11,I12}. As part of this
approach, a version of GR called the symmetric teleparallel equivalent of GR 
$\left( STEGR\right) $ was suggested \cite%
{I13,I14,I15,I16,I161,I16,I161,I162,I163,I164,I165,I166,I167,I168,I169}.
This model utilizes a non-metric connection for spacetime geometry, but with
the constraint that the total curvature and torsion are both zero. STEGR can
also be extended to $f(Q)$ gravity \cite{I17}, which does not require the
Equivalence Principle as an initial assumption. This theory can be
approached using gauge theories and has several benefits, including no
significant coupling issues caused by extra scalar modes \cite{I18}.
Recently, $f(Q)$ gravity has gained attention \cite{I19,I20}, particularly
in regards to explaining problems with late-time acceleration and dark
energy \cite{I21,I22,I23}. However, its effects on the early universe have
not been fully explored.

Following inflation, there is a period during which the energy density
stored in the inflaton converts into a thermal bath consisting of
relativistic particles \cite{I24,I25,I26,I27,I28,I29,I30,I31}. This process
is known as reheating and takes place between the end of inflation and the
start of the radiation-dominated epoch. During the reheating process,
ordinary matter is produced as a result of the inflaton field losing energy.
In the simplest case, reheating happens through the inflaton's perturbative
decay into particles of standard model matter as it oscillates around the
minimum of its potential \cite{I40,I50}. However, this scenario has been
challenged due to its inability to explain the cohesive nature of the
inflaton field \cite{I60,I70}. In some instances, a preheating stage takes
place before reheating, during which particles are created through
nonperturbative mechanisms like parametric resonance decay \cite{I80}. The
potential for both preheating and reheating to occur between the end of
inflation and the start of the radiation-dominated epoch was explored in 
\cite{I27a}.

The structure of our paper is outlined as follows: Section \ref{sec2}
provides an overview of the fundamental principles of $f(Q)$ gravity.
Section \ref{sec3} introduces the tree level inflation model, while Section %
\ref{sec4} describes the tree level inflationary scenario within $f(Q)$
gravity. In Section \ref{sec5}, we summarize our findings and present our
conclusions.

\section{Inflationary cosmology in $f(Q)$ gravity}

\label{sec2}

In Weyl-Cartan geometry, gravitational effects are caused not only by a
change in the direction of a vector in parallel transport, but also by a
change in its length. Non-metricity represents the geometric variation in
the length of a vector and is theoretically defined as the covariant
derivative of the metric tensor. In this paper, we will look at the modified
Einstein-Hilbert action in f(Q) symmetric teleparallel gravity, which is
written as,%
\begin{equation}
S=\int \frac{1}{2\kappa }f(Q)\sqrt{-g}d^{4}x+S_{\phi },  \label{eqn1}
\end{equation}%
where $f(Q)$ is a generic function of the non-metricity scalar $Q$, $g$ is
the determinant of the metric tensor $g_{\mu \nu }$ i.e. $g=\det \left(
g_{\mu \nu }\right) $,\ $\kappa =1/M_{p}^{2}$\ and $S_{\phi }$ is the
inflaton action defined in the next section. In addition, the non-metricity
scalar $Q$ is given as,%
\begin{equation}
Q\equiv -g^{\mu \nu }(L_{\,\,\,\alpha \mu }^{\beta }L_{\,\,\,\nu \beta
}^{\alpha }-L_{\,\,\,\alpha \beta }^{\beta }L_{\,\,\,\mu \nu }^{\alpha }).
\label{eqn2}
\end{equation}

We obtain the field equations of $f(Q)$ gravity by varying the gravitational
action (\ref{eqn1}) with regard to the metric tensor $g_{\mu \nu }$, 
\begin{equation}
-\frac{2}{\sqrt{-g}}\nabla _{\beta }\left( f_{Q}\sqrt{-g}P_{\,\,\,\,\mu \nu
}^{\beta }\right) -\frac{1}{2}fg_{\mu \nu }-f_{Q}(P_{\mu \beta \alpha
}Q_{\nu }^{\,\,\,\beta \alpha }-2Q_{\,\,\,\mu }^{\beta \alpha }P_{\beta
\alpha \nu })=\kappa T_{\mu \nu }.  \label{eqn3}
\end{equation}%
Here, $f_{Q}=\frac{df\left( Q\right) }{dQ}$, and $\nabla _{\beta }$ is the
covariant derivative is shown. We now investigate a homogeneous and
spatially flat Friedmann-Lemaitre-Robertson-Walker (FLRW) metric. As a
result, the analogous non-metricity scalar is $Q=6H^{2}$. In our current
investigation, we suppose that the Universe is a perfect non-viscous fluid
with the energy-momentum tensor given by%
\begin{equation}
T_{\nu }^{\mu }=diag\left( -\rho ,p,p,p\right) ,  \label{eqn4}
\end{equation}%
The updated Friedmann equations are stated as where $p$ is the ideal
non-viscosity fluid pressure and the Universe's energy density is $\rho $. 
\begin{equation}
\kappa \rho =\frac{f}{2}-6FH^{2}  \label{eqn5}
\end{equation}%
and%
\begin{equation}
\kappa p=-\frac{f}{2}+6FH^{2}+2\left( \dot{F}H+F\dot{H}\right) .
\label{eqn6}
\end{equation}%
In this case, $\left( \text{\textperiodcentered }\right) $ denotes a
derivative with regard to cosmic time $\left( t\right) $, while $F\equiv
f_{Q}$ denotes differentiation with respect to $Q$. Combining Eqs (\ref{eqn5}%
) and (\ref{eqn6}) yields the evolution equation for the Hubble parameter $H$
as,%
\begin{equation}
\overset{.}{H}+\frac{\overset{.}{F}}{F}H=\frac{\kappa }{2F}\left( \rho
+p\right) .  \label{eqn7}
\end{equation}%
The field equations of Einstein (\ref{eqn5}) and (\ref{eqn6}) can be thought
of as extended symmetric teleparallel counterparts to conventional
Friedmann's equations, with added terms from space-time non-metricity and
the trace of the energy-momentum tensor $T$ acting as an effective
component. As a result, the effective energy density $\rho _{eff}$ and
effective pressure $p_{eff}$ are calculated.%
\begin{equation}
3H^{2}=\kappa \rho _{eff}=\frac{f}{4F}-\frac{\kappa }{2F}\rho ,  \label{eqn8}
\end{equation}%
\bigskip with%
\begin{equation}
2\dot{H}+3H^{2}=-\kappa p_{eff}=\frac{f}{4F}-\frac{2\dot{F}H}{F}+\frac{%
\kappa }{2F}\left( \rho +2p\right) .  \label{eqn9}
\end{equation}%
Taking into account Eqs. (\ref{eqn5}) and (\ref{eqn6}) one gets%
\begin{equation}
\rho =\frac{f-12H^{2}F}{2\kappa }.  \label{eqn10}
\end{equation}%
The inflationary universe model was developed to address a number of
cosmological concerns, including the horizon, flatness, and monopole issues.
The Klein-Gordon equation $\ddot{\phi}+3H\dot{\phi}+V^{\prime }=0$, which
governs the dynamics of the scalar field $\phi $, connects the scalar field
with the scalar potential $V(\phi )$ in inflation theory. The slow-roll
approximation involves ignoring the kinetic components and the second
temporal derivatives of the field $\phi $, i.e. $\dot{\phi}^{2}\ll V$ and $%
\ddot{\phi}\ll V$ respectively, thus,%
\begin{eqnarray}
H^{2} &\sim &\frac{V(\phi )}{3M_{p}^{2}},  \label{eqn11} \\
\dot{\phi} &\sim &-\frac{V^{\prime }\left( \phi \right) }{3H}.  \label{eqn12}
\end{eqnarray}%
The Hubble parameter may be used to generate the slow-roll parameters as 
\cite{A1}%
\begin{eqnarray}
\epsilon &=&-\frac{\dot{H}}{H^{2}},  \label{eps13} \\
\eta &=&\frac{\dot{\epsilon}}{H\epsilon }\approx -\frac{\ddot{H}}{2\dot{H}H}.
\label{eps14}
\end{eqnarray}%
we define the number of e-folds associated with inflation from the beginning 
$t$ to $t_{end}$\ as follows 
\begin{equation}
N\equiv \ln \left( \frac{a_{end}}{a}\right) =\int_{t}^{t_{end}}Hdt,
\label{eqn15}
\end{equation}

The spectral index $n_{s}$ describes the relationship between curvature
disturbances and slow roll parameters. The e-folding number $N$, which
reflects the pace of the Univers expansion during this era, the tensor to
scalar perturbations ratio $r$, and the spectral index $n_{s}$ are also used
to investigate the period of inflation.%
\begin{eqnarray}
n_{s}-1 &=&-6\epsilon +2\eta ,  \label{eqn16} \\
r &=&16\epsilon .  \label{eqn17}
\end{eqnarray}%
Next, we will discuss how the tree level potential inflationary model can be
described in the context of $f(Q)$ gravity.\ \ 

\section{The tree level potential in $f(Q)$ gravity}

\label{sec3}

\subsection{The inflationary parameters on $f(Q)$ gravity}

In the context of cosmic inflation, we consider the model $f(Q)=\alpha Q,$
with $\alpha =F\neq 0$, by taking Eqs.\ (\ref{eqn7}), (\ref{eqn8}), (\ref%
{eqn9}) and (\ref{eqn10})\ we obtain the following equations%
\begin{eqnarray}
\dot{H} &=&\frac{\kappa \rho \left( 1+\omega \right) }{2\alpha },  \label{Hp}
\\
3H^{2} &=&\kappa \rho _{eff}=-\frac{\kappa \rho }{\alpha },  \label{eqnH} \\
2\dot{H}+3H^{2} &=&-\kappa p_{eff}=\frac{\kappa \omega \rho }{\alpha },
\label{eqnHH} \\
\rho &=&-\frac{3\alpha H^{2}}{\kappa }.
\end{eqnarray}%
\newline
In of $f(Q)$ gravity, we consider the tachyonic field leads from Eqs. (\ref%
{eqn20}), (\ref{eqn21}) along with Eqs.(\ref{eqn17}), (\ref{eqn18}) gives%
\begin{eqnarray}
3H^{2} &=&-\frac{\kappa }{\alpha }V(\phi ),  \label{pp} \\
2\dot{H}+3H^{2} &=&\frac{\kappa }{\alpha }\omega V(\phi ).  \label{p}
\end{eqnarray}%
Taking into account the previous two equations one gets%
\begin{equation}
\dot{H}=\frac{\kappa }{2\alpha }\dot{\phi}^{2},  \label{H.}
\end{equation}%
after the derivation of Eq.(\ref{eqnH}) and using Eq. (\ref{eqnHH}), we
obtain the modified equation of motion%
\begin{equation}
\ddot{\phi}+3H\dot{\phi}+V^{\prime }=0.
\end{equation}%
From Eqs. (\ref{eps2}) we know that the condition $\eta \ll 1$ leads to $%
\ddot{\phi}\ll 3H\dot{\phi},$\ which means that the time derivative of the
tachyonic field is given as $\dot{\phi}\sim -V^{\prime }\left( \phi \right)
/3H$. Now, let us use the slow-roll parameters that will be calculated in
the context of the $f(Q)$ gravity. For this purpose, we should insert Eqs. (%
\ref{pp}) and (\ref{H.}) into the slow roll parameter, knowing that we need
to consider the case $\dot{\phi}^{2}<2/3$ which gives%
\begin{eqnarray}
\epsilon &\approx &\frac{3}{2}\frac{\dot{\phi}^{2}}{V}\simeq -\frac{\alpha }{%
2\kappa }\left( \frac{V^{\prime }}{V}\right) ^{2}, \\
\eta &\approx &-\frac{\ddot{\phi}}{H\dot{\phi}}\simeq -\frac{\alpha }{\kappa 
}\frac{V^{\prime \prime }}{V},
\end{eqnarray}%
the e-folds number for this chosen model can be written using Eq.(\ref{pp})
and Eq.(\ref{eqn12}) in the following way{}%
\begin{equation}
N\equiv \int_{\phi _{k}}^{\phi _{end}}\frac{H}{\dot{\phi}}d\phi \approx
-\int_{\phi _{end}}^{\phi _{k}}\frac{\kappa }{\alpha }\frac{V}{V^{\prime }}%
d\phi .  \label{N}
\end{equation}%
Our goal next is to calculate the inflationary parameters for tree level
potential in the chosen model of $f(Q)$ gravity.

\subsection{ Constraints on the tree level potential in $f(Q)$ gravity}

The proposed tree level Higgs potential \cite{I4}, \cite{I5}, \cite{A2} is
given by 
\begin{equation}
V(\phi )=V_{0}\left[ 1-\left( \frac{\phi }{\mu }\right) ^{2}\right] ^{2},
\label{eqn18}
\end{equation}

here $V_{0}$ is the vacuum energy density at the origin and $\mu $ is the
vacuum expectation value of the inflation, this potential will be discussed
for three branches: $\phi ^{2}\gg \mu ^{2}$ (\textit{AV branch}), $\phi \sim
\mu $ (\textit{Near branch}), and $\phi ^{2}\ll \mu ^{2}$ (\textit{BV branch}%
).

\subsubsection*{\textit{AV branch}: $\protect\phi ^{2}\gg \protect\mu ^{2}$}

In the case of an \textit{AV branch}, the potential becomes%
\begin{equation}
V(\phi )=V_{0}\left( \frac{\phi }{\mu }\right) ^{4},  \label{eqn19}
\end{equation}%
the slow-roll parameters in this case are determined by the values of the
field at the beginning of inflation $\phi _{k}$ and is\ given by%
\begin{eqnarray}
\epsilon &\simeq &-\frac{8\alpha }{\kappa }\frac{1}{\phi _{k}^{2}},
\label{eqn20} \\
\eta &\simeq &-\frac{12\alpha }{\kappa }\frac{1}{\phi _{k}^{2}}.
\label{eqn21}
\end{eqnarray}%
Equation (\ref{N}), on the other hand, offers the integral growth from $\phi
_{k}$ to $\phi _{end}$ as%
\begin{equation}
N\simeq -\frac{\kappa }{8\alpha }\left( \phi _{k}^{2}-\phi _{end}^{2}\right)
\simeq -\frac{\kappa }{8\alpha }\phi _{k}^{2}.
\end{equation}

Now since the inflationary parameters are determined by the value of the
inflaton field $\phi _{k}$, we can link these parameters to observation
using Eq.(\ref{eqn16}) to obtain $\left( n_{s}-1\right) /3=8\alpha /\kappa
\phi _{k}^{2}.$

\subsubsection*{\textit{Near branch}: $\protect\phi \sim \protect\mu $}

We will analyze the physical consequences of the solutions in the situation
of $\phi \sim \mu $ when the tree level potential becomes%
\begin{equation}
V(\phi )=V_{0}\left[ 1-\left( \frac{\phi }{\mu }\right) ^{2}\right] ^{2}=%
\frac{V_{0}}{\mu ^{4}}\left( \phi +\mu \right) ^{2}\left( \mu -\phi \right)
^{2}  \label{eqn23}
\end{equation}%
in the interests of clarity, we take the approximation: $\phi +\mu =2\mu $
where $\phi \sim \mu $, whereas the potential becomes%
\begin{equation}
V(\phi )\simeq \frac{4V_{0}}{\mu ^{2}}\left( \mu -\phi \right) ^{2}.
\label{eqn24}
\end{equation}

In this context, one can also consider the slow-roll parameters for this
potential by%
\begin{eqnarray}
\epsilon &\simeq &-\frac{2\alpha }{\kappa }\frac{1}{\left( \phi _{k}-\mu
\right) ^{2}},  \label{eqn25} \\
\eta &\simeq &-\frac{2\alpha }{\kappa }\frac{1}{\left( \phi _{k}-\mu \right)
^{2}}.  \label{eqn26}
\end{eqnarray}

On the other hand, for the\textit{\ Near branch}\emph{,} the number of
e-folds gives 
\begin{equation}
N\simeq -\frac{\kappa }{2\alpha }\left[ \left( \phi _{k}-\mu \right)
^{2}-\left( \phi _{end}-\mu \right) ^{2}\right] \simeq -\frac{\kappa }{%
2\alpha }\left( \phi _{k}-\mu \right) ^{2}  \label{eqn27}
\end{equation}%
Because the inflationary parameters are defined by the value of the inflaton
field $\phi _{k}$, we can now relate these parameters to observations to
obtain $\left( n_{s}-1\right) =8\alpha /\kappa \left( \phi _{k}-\mu \right)
^{2}.$

\subsubsection*{\textit{BV branch}: $\protect\phi ^{2}\ll \protect\mu ^{2}$}

In this section, we will focus on inflationary potential, which may be
written in the \textit{BV branch} as \cite{I5}%
\begin{equation}
V(\phi )=V_{0}\left( 1-2\frac{\phi ^{2}}{\mu ^{2}}\right) .  \label{eqn1a}
\end{equation}%
In this situation, the slow-roll settings can also be considered. knowing
that $\left( \phi /\mu \right) ^{2}\ll 1$. The two first parameters for this
branch are given by%
\begin{eqnarray}
\epsilon &\simeq &-\frac{8\alpha }{\kappa }\frac{\phi _{k}^{2}}{\mu ^{4}},
\label{eqn2a} \\
\eta &\simeq &\frac{4\alpha }{\kappa }\frac{1}{\mu ^{2}}.  \label{eqn3a}
\end{eqnarray}%
On the other hand, the number of e-folding $N$ can be expressed as%
\begin{equation}
N\simeq -\frac{\kappa }{4\alpha }\mu ^{2}\left( \ln \left( \phi
_{end}\right) -\ln \left( \phi _{k}\right) \right) ,  \label{eqn4a}
\end{equation}%
here we can simply calculate $\phi _{end}$\ considering $\epsilon =1$\ that
can be applied at the end of inflation when $\phi \rightarrow \phi _{end}.$\ 

We can now link the inflationary parameters to observations. However, this
time we have to specify precise values of $\phi _{k},$\ $\mu ,$\ and $\alpha 
$\ parameters since the spectral index parameter is not specified by the
value of the inflaton field $\phi _{k}$ and taking into account the
approximation $\left( \phi /\mu \right) ^{2}\ll 1,$ $n_{s}$ is given by the
following equation$\ \left( n_{s}-1\right) =8\alpha /\kappa \mu ^{2}.$

\section{Reheating mechanism from the inflaton decay}

\label{sec4}

The process of reheating plays a crucial role in the evolution of the
universe. It marks the emergence of matter towards the end of the rapid
expansion phase when the inflaton field breaks down, and standard model
particles are formed, resulting in a rise in temperature that brings about a
restoration of the standard Big Bang cosmology model. \cite{A3}. As the
universe expands, the amplitude of the oscillations gradually decreases, and
the oscillating field transfers its energy to the newly formed particles. In
addition, we take into account the crucial interaction Lagrangian \cite{A4}.%
\begin{equation}
L_{int}=\frac{1}{2}g^{2}\phi ^{2}\chi ^{2}+h\psi \bar{\psi}\phi ,
\label{eqn28}
\end{equation}%
where $h$ and $\sigma $\ are dimensionless couplings and $\sigma $ has the
dimensions of mass. The decay rate of the inflaton field is defined in the
following way $\Gamma =\Gamma _{\left( \phi \rightarrow \chi \chi \right)
}+\Gamma _{\left( \phi \rightarrow \psi \psi \right) }$ with $\Gamma
_{\left( \phi \rightarrow \chi \chi \right) }=g^{4}\sigma ^{2}/8\pi m_{\phi
},$\ $\Gamma _{\left( \phi \rightarrow \psi \psi \right) }=h^{2}m_{\phi
}/8\pi $ and $m_{\phi }$\ is the mass of the inflaton field.\ 

Next, we demonstrate the correlation between the decay rate and reheating
parameters. We start with the Friedmann equation in the $f(Q)$ gravity
scenario, which is $H%
%TCIMACRO{\U{b2}}%
%BeginExpansion
{{}^2}%
%EndExpansion
=-\kappa \rho /3\alpha $. Subsequently, we express the energy density at the
onset of the reheating phase as a function of the reheating temperature.%
\begin{equation}
\rho _{re}=\frac{\pi ^{2}}{30}g_{\ast }T_{re}^{4},  \label{eqn29}
\end{equation}%
The reheating relativistic degree of freedom is denoted as $g_{\ast }$. In
the reheating phase, we can substitute $\rho $ with $\rho _{re}$ to obtain
the Friedmann equation expressed as follows:%
\begin{equation}
H^{2}=-\frac{\kappa }{3\alpha }\frac{\pi ^{2}}{30}g_{\ast }T_{re}^{4}
\label{eqn30}
\end{equation}

After inflation ends, the energy densities of the scalar field and radiation
become comparable when $H\simeq \Gamma $. From this point onwards, the
Universe is governed by radiation, which determines the reheating
temperature. The temperature can be expressed as:%
\begin{equation}
T_{re}=\left( -\frac{3\cdot 30\alpha }{\pi ^{2}g_{\ast }}\frac{\Gamma ^{2}}{%
\kappa }\right) ^{\frac{1}{4}}.  \label{eqn31}
\end{equation}%
\begin{figure}[H]
\centering
\includegraphics[width=15cm]{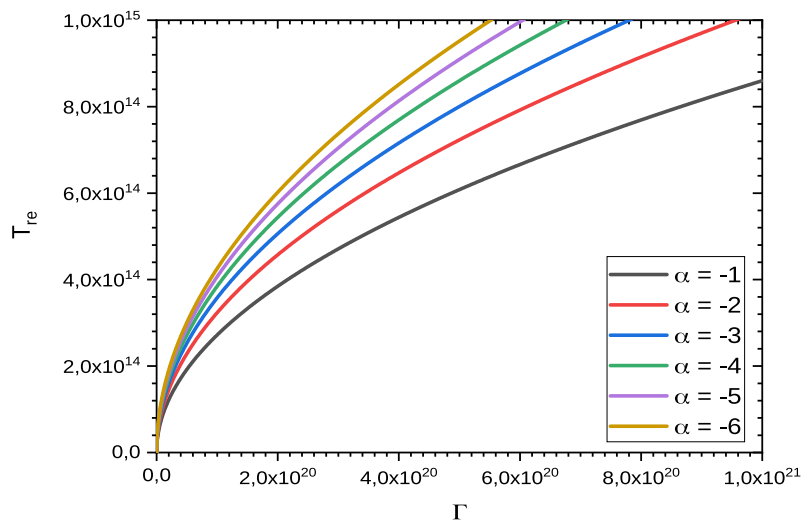}
\caption{The evolution of the reheating temperature as a function of the
total decay rate for different values of $\protect\alpha .$}
\label{fig:3}
\end{figure}

In Fig. (\ref{fig:3}) we plot the variation of the reheating temperature $%
T_{re}$ as a function of the $\alpha $\ parameter related to the $f(Q)$
gravity model, for mathematical consistency we observe that our model is
consistent only for negative values of $\alpha .$ To study the behavior of $%
T_{re}$\ we chose different values of $\alpha $ bounded in the following way 
$\alpha \leq -1$\ which makes the increasing curve of reheating temperature
moves towards lower values of reheating temperature for $\Gamma \propto
10^{21}GeV$\ , our results illustrate the predictions of Eq.(\ref{eqn31})\
which lead to the conclusion that higher values of the decay rate energy
will lead to higher reheating temperature.

\section{Conclusion}

\label{sec5}

In recent years, despite the undeniable success of 'the standard model of
cosmology' governed by Einstein's general theory of relativity (GTR), its
limitations have become apparent in various aspects. These include
cosmological singularity and black holes, tensions in $H_{0}$ and $\sigma
_{8}$ on observational grounds, and most importantly, the need for yet
undetected dark sectors to explain the late-time cosmic acceleration. These
limitations have prompted researchers to explore alternative theories of
gravity. Teleparallel gravity has emerged as a significant area of research
in this regard. Specifically, researchers have considered replacing the
Levi-Civita connection, which forms the foundation of GTR, with an affine
connection on spatially flat spacetime with vanishing torsion. This allows
non-metricity to entirely define gravity, giving rise to the theory known as
'symmetric teleparallel gravity'. The focus of our paper is to examine tree
level Higgs inflation and reheating in $f(Q)$ gravity. Our study involves
limiting the inflationary parameters and the reheating temperature. We begin
by outlining the formalism for tree level inflation within the chosen
modified $f(Q)$ model. We discovered that the slow roll and e-folds number
related to inflation relies on the $f(Q)$ model $\alpha $ parameter.
Additionally, we have computed the reheating in relation to the decay rate.
We selected various values of $\alpha $ that are bounded by $\alpha \leq -1$%
. This constraint causes the upward trend of the reheating temperature curve
to shift towards lower values of reheating temperature. This leads to the
conclusion that increasing the energy of the decay rate results in a
corresponding increase in the reheating temperature.

\end{document}